\documentclass[conference]{IEEEtran}
\IEEEoverridecommandlockouts
\usepackage{amsmath,amssymb,enumerate}
\usepackage{cite,graphics,graphicx}
\usepackage{theorem}

{\theorembodyfont{\rmfamily}
   \newtheorem{Exa}{{\textbf Example}}[section]}

\begin{document}
\title{AONT-LT: a Data Protection Scheme for Cloud and Cooperative Storage Systems}

\author{\IEEEauthorblockN{M. Baldi, N. Maturo, E. Montali, F. Chiaraluce}
\IEEEauthorblockA{Dipartimento di Ingegneria dell'Informazione (DII)\\
Universit\`a Politecnica delle Marche\\
Ancona, Italy\\
$\lbrace m.baldi, n.maturo\rbrace @univpm.it, eugenio.montali@gmail.com, f.chiaraluce@univpm.it$
}~\thanks{This work was supported in part by the MIUR project “ESCAPADE” (Grant
RBFR105NLC) under the “FIRB - Futuro in Ricerca 2010” funding program.}}

\maketitle

\begin{abstract}
%\boldmath
We propose a variant of the well-known AONT-RS scheme for dispersed storage systems. The novelty consists in replacing the Reed-Solomon code with rateless Luby transform codes. The resulting system, named AONT-LT, is able to improve the performance by dispersing the data over an arbitrarily large number of storage nodes while ensuring limited complexity. The proposed solution is particularly suitable in the case of cooperative storage systems. It is shown that while the AONT-RS scheme requires the adoption of fragmentation for achieving widespread distribution, thus penalizing the performance, the new AONT-LT scheme can exploit variable length codes which allow to achieve very good performance and scalability.
\end{abstract}
\vspace{0.1in}

% \fi
\IEEEpeerreviewmaketitle

\section{INTRODUCTION}
\label{Intro}
In the last few years, online storage services have obtained an increasing commercial success.  Solutions like iCloud, Dropbox, Skydrive, etc., are well known and used on a daily basis by millions of people all over the world. These solutions are based on the same approach: the service provider buys or rents a large number of servers in which authorized clients are able to store their data. By exploiting such a large number of servers, information protection can be achieved through suitable data encoding and slicing. In this case, the client or provider software decomposes the file into smaller blocks (from now on called \textit{slices}) which are properly processed and then dispersed to the storage sites. Dually, when a client desires to read a file, he retrieves some subset of the slices, which are combined to reconstruct the original file. Obviously, a structure of this type implies very high initial costs, necessary to guarantee a high performance service in terms of storage availability and connection capabilities; moreover, a crucial issue concerns reliability against failures.

To implement this kind of services with lower costs, a new approach has recently emerged and is now rapidly growing: it consists in using cooperative storage, which means that the storage capacity is provided directly by the clients themselves who, co-operating in the cloud, make their own storage facilities available to the others. This approach offers some evident advantages: first of all the service provider only needs a low number of servers, acting as coordinators for the access to the service. Secondly, increasing the number of users yields an increase of the storage capabilities.

So, in a cooperative storage system (CSS), even more than in a conventional distributed storage system (DSS), it is expected that the number of storage nodes becomes very large, offering the possibility of dispersing the data over a huge number of distinct nodes.

Actually, data slicing is at the basis of a security paradigm that, over the cloud, does not need the use of pre-shared encryption keys. In such a scheme, the storage system transforms a file into $n$ distinct slices and only the client (or attacker) having access to, at least, $k$ out of the $n$ slices is able to reconstruct the file. Instead, the client or attacker with a number of slices lower than $k$ is not able to get any information. This basic principle has been applied since the pioneer works by Shamir \cite{Shamir1979} and Rabin \cite{Rabin1989}, and subsequently confirmed by McEliece and Sarwate \cite{McEliece1981} who disclosed the relationship with Reed-Solomon (RS) coding schemes. Actually, the strength of RS codes in this kind of applications is in the fact that these codes are maximum distance separable (MDS), which implies it is sure that by knowing $k$ slices (arbitrarily distributed among the $n$) the whole file can be successfully recovered. Such a feature implies a characteristic step function behavior for the curve of successful decoding probability as a function of the erasure rate $p_e$ (defined as the percentage of storage nodes that cannot be accessed, by the client or the attacker), which is $1$ for $p_e \leq k/n$ and $0$ for $p_e > k/n$. Obviously we expect that the value of $p_e$ for the attacker is greater than that for the client.

It is intuitively reasonable that by increasing the number of slices (and then the number of storage nodes if, for example, each node stores a single slice) we can have some important advantages. In particular, a larger number of slices can result in:
\begin{itemize}
\item[-] the possibility, for the client, to tolerate higher values of $p_e$ without incurring in information losses;
\item[-] the need, for the attacker, to crack a larger number of nodes to steal the data;
\item[-] the opportunity to exploit even minimal storage units, as offered by service users, to store a significant piece of information.
\end{itemize}

Unfortunately, however, when using an RS($n, k$) code the number of slices is related to the size of the Galois Field (GF) where the RS code is defined: the greater the value of $n$ the larger the GF. As it will be explained in Section \ref{AONT_RS}, this has an impact on the complexity: no proposals have appeared till now using RS codes defined over Galois fields with more than 256 elements, that is $n \le 255$, and this severely limits the achievable slicing level. To overcome the problem, different coding schemes have been adopted as well. Among them, a smart solution consists in using Luby transform (LT) codes \cite{Cao2012}.

LT codes, proposed by Michael Luby in 2002 \cite{Luby2002}, are an example of universal erasure correcting codes. The length of an LT codeword can be freely varied in real time during the encoding operations; for this reason, LT codes are also said to be rateless: in fact, while the input word length ($k$) is fixed, the output word length ($n$) can be chosen ``on the fly''. It is easy to understand that such a feature makes LT codes very interesting in cooperative storage scenarios. Using LT codes, in fact, there is no limit, in principle, on the value of $n$. Moreover, the coding and decoding operations have low complexity (much smaller than those of the RS codes), since they are based on simple XOR operations and efficient belief propagation (BP), respectively. On the other hand, an LT code is not MDS and, in this sense, it exhibits a sub-optimal behavior: to recover the whole file, it is necessary to have access to a number of slices (slightly) larger than $k$ and, generally speaking, the curve of the successful decoding probability is not the step function typical of the RS codes. Though important, these drawbacks can be largely mitigated through a suitable code design and, as a matter of fact, LT codes appear as a valid alternative to RS codes also from the data repair and data retrieval standpoints \cite{Cao2012}.

In this paper, we exploit LT codes to propose a new secure and reliable information processing algorithm for CSS, which is based on the combination of these codes with an all-or-nothing-transform (AONT) \cite{Rivest1997b}. The security challenge is particularly evident in the case of CSS \cite{Mulazzani2011}, \cite{Drago2012}. This is because, in a CSS, data are not delivered to a trusted entity that controls the storage nodes and guarantees their trustworthiness; hence, the nodes are intrinsically vulnerable and exposed to the action of malicious eavesdroppers. The AONT ensures the confusion and diffusion properties that are at the basis of any efficient secure system, while the combination with an error correcting code allows to make the system robust against some node failures. The AONT has been combined with systematic RS codes in \cite{Resch2011}, showing its ability to achieve improved computational performance, security and integrity. The AONT-RS system, however, suffers the limitation mentioned above on the maximum number of usable storage nodes.  A conceptually simple way to overcome the limit consists in applying RS coding to single fragments of the input file. We call this scheme AONT-RS with fragmentation. As will be shown afterward, however, fragmentation penalizes the performance and the step behavior of the successful decoding probability may be lost. So, in this paper we propose, for the first time to the best of our knowledge, an AONT-LT system and we prove, through explicit examples, that it is able to outperform the AONT-RS with fragmentation.

The organization of the rest of the paper is as follows. In Section \ref{AONT_RS} we shortly describe the AONT-RS system and introduce the fragmentation option; in Section \ref{AONT_LTS} we present the new AONT-LT proposal; in Section \ref{Performance} we discuss some numerical results. Finally, Section \ref{Concl} concludes the paper.

\section{AONT-RS}
\label{AONT_RS}

In principle, the security of a storage system could be achieved by using a classic symmetric encryption scheme. It is easy to understand, however, that such a conceptually simple procedure can put serious problems as regards the key management and distribution. If the same password used to access the system is employed as a cryptographic key (that could seem an easy-to-implement solution), the loss of the password, that is not dramatic from the access standpoint as the system is able to generate a new password, is disruptive from the encryption standpoint as the new password is obviously unable to decrypt the files encrypted with the old, lost password. Hence more complex, and dangerous mechanisms should be conceived for key management.

A desirable property is to avoid the secure storage of encryption keys, but it is not foregone that this is achieved with low computational and storage costs. As mentioned in Section \ref{Intro}, a smart technique for ensuring security and reliability in a DSS without the need of external keys consists in using AONT-RS \cite{Resch2011}. The popular Symform service implements an instance of the AONT-RS, by using AES-256 as the basis for the AONT and the RS(96, 64) for coding.

\begin{figure}[tb]
\centering
\includegraphics[width=3.5in]{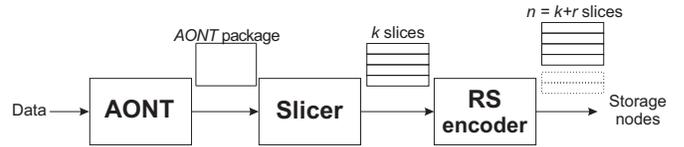}
\caption{AONT-RS principle.}
\label{fig.AONT_RS}
\end{figure}

The basic elements of the AONT-RS scheme are shown in Fig. \ref{fig.AONT_RS}. This is a highly simplified figure, since it contains the only information necessary to understand the variants introduced by the new proposal in Section \ref{AONT_LTS}. Further details, here omitted for the sake of brevity, can be found in \cite{Resch2011}.

The data entering the AONT are composed of $s + 1$ words $d_0, . . . , d_s$\footnote{$d_0, . . . , d_{s-1}$ are data words while $d_s$ is an extra word, called canary, which allows to check the integrity of the data after decoding.}; a random key $K$ is chosen and a symmetric key-based encryption algorithm $E$, like AES, is used to obtain
\begin{equation}
c_i = d_i \oplus E(K, i+1).
\label{eq.AONT}
\end{equation}

A final block is then computed as $c_{s + 1} = K \oplus h$, where $h$ is a hash digest of the $s + 1$ codewords $c_i$, obtained by using a standard hash algorithm such as SHA-256 (the hash output is at least as long as $K$). Coherent with the previous considerations, it is important to underline that $K$ is an internal key necessary to feed the AONT, but does not represent a secret to be shared among authorized users. The AONT package, formed by a set of $s$ AONT words and completed with padding bits if necessary, is sent to a slicer that forms groups of $k$ slices; the latter are then encoded by the RS($n, k$) code, thus obtaining, at the output, $n$ slices. Each slice is eventually stored in a different node. To retrieve the data, a client (or an attacker) must recover a minimum of $k$ slices. After RS decoding, the AONT must be reversed as well, but this is trivial starting from the entire package since, by using simple XOR operations, the random key $K$ can be reconstructed and then used to decrypt the encrypted data. Finally, the canary is checked to detect possible corruption. A more comprehensive description can be found in \cite{Resch2011}.

It is well known that the value of $n$ for an RS code is determined by the size $q = 2^m$ of the Galois field GF($q$) where the code symbols are defined. More precisely, we have $n \leq q - 1$. Large values of $n$ reflect into a greater size of the GF, and this has an impact on the complexity, which becomes higher and higher. For this reason, a common value for $q$ is $256$, which has also the practical advantage to represent each element of the field (symbol) as a byte ($m = 8$).
%It must be noted that $m$ is also the number of bits in each slice.
Larger values of $q$ are difficult to sustain as they become computationally intractable. According to this choice, $255$ is also the maximum number of storage nodes in the CSS, under the hypothesis of uploading only $1$ slice per node. Wishing to adopt an RS code with rate $R_c \approx 1/2$, that is another common choice as well, this means that the slicer must divide the data into groups of $128$ slices, with the corresponding slicing level.

Wishing to store a file in a number of nodes $n_s > 255$ and avoiding, for the reasons explained, the adoption of a GF with greater size, a possible solution is as follows.

\begin{figure}[tb]
\centering
\includegraphics[width=3.5in]{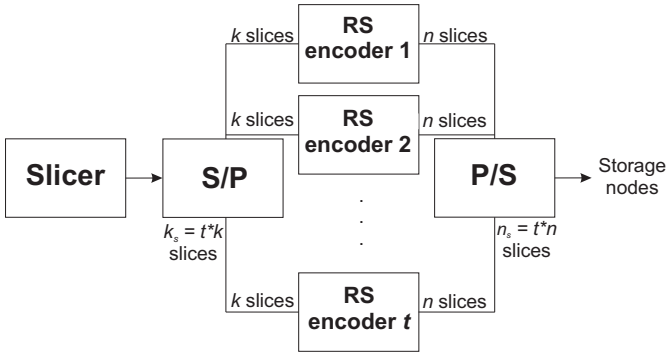}
\caption{AONT-RS with fragmentation. S/P and P/S stay for slice series-parallel and parallel-series conversion respectively.}
\label{fig.AONT_tRS}
\end{figure}

According to the chosen value of $n_s$, we have $k_s = R_c \times n_s$ slices at the RS encoder input. These $k_s$ slices are split into $t$ fragments of $k$ slices each, and any block is encoded with a separate RS code with $n \leq 255$. The situation is schematically shown in Fig. \ref{fig.AONT_tRS}. So, at the output of the parallel of $t$ RS encoders we have $n_s$ slices as desired. We will refer to this solution as AONT-RS with fragmentation.

The scheme in Fig. \ref{fig.AONT_tRS}, that replaces the right part of Fig. \ref{fig.AONT_RS}, implicitly assumes that the AONT is applied to the whole block of $k_s$ slices. This implies that, for correctly inverting the AONT function, all the $t$ RS decoding operations must be successful. This fact may have an impact on the performance, as clarified by the following simple example.

\begin{Exa}
\label{Ex:one}
The channel model suited for the considered scenario is the binary erasure channel (BEC). Let us consider a system with $n_s = 765$ nodes available; this implies that $t = 3$ RS($255, 127$) codes are used in parallel. Let us suppose that $p_e = 0.4$ and that, incidentally, the first $306$ slices (i.e, the nodes where they are stored) are inaccessible, for example as damaged. The situation is depicted in Fig. \ref{fig.Example}. In this case, the number of slices available for the three decoders is very different: $0$ for the first one, $204$ for the second and $255$ for the third. So, the first decoder will fail, while the second and the third decoders will experience a successful decoding. As the AONT is applied on the whole set of slices, however, one decoding failure is enough to prevent the correct reversal of the AONT, so that the original file cannot be retrieved. It is important to observe that this occurs for a value of $p_e$ smaller than the code rate ($\approx 0.5$), while in case of using a single code the same failure probability would not produce any information loss. Clearly, if the $306$ erasures were differently distributed (for example, $102$ for each codeword) all decoders could be successful.
\begin{figure}[tb]
\centering
\includegraphics[width=3.5in]{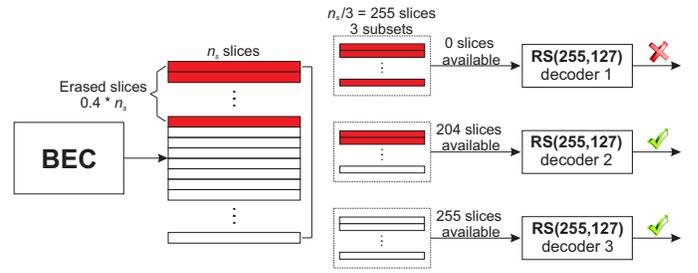}
\caption{Example of decoding of AONT-RS with fragmentation in the presence of failures ($p_e = 0.4$).}
\label{fig.Example}
\end{figure}
\end{Exa}

Example \ref{Ex:one} shows that when the AONT-RS system with fragmentation is used and the AONT is applied to the whole file, the successful decoding probability as a function of the failure probability may degrade; in practice, all fragments must be successfully decoded in order to be able to reverse the AONT and this implies that the curve has no longer the step behavior which is typical of a single RS code. Numerical examples will be shown in Section \ref{Performance}. Symmetrically, the attacker has normally access to a smaller number of nodes, which is equivalent to assume for him a larger $p_e$. Because of the lack of the step behavior, his successful decoding probability may be greater than $0$ even for $p_e > R_c$.

As an alternative option, one can think to apply the AONT not to the whole block of $k_s$ slices but individually to every block of $k$ slices entering each encoder. However, this solution is clearly more insecure, as it would permit an attacker to disclose significant portions of the information, even with rather high values of $p_e$. In fact, in this case it is sufficient that decoding of a single fragment succeeds in order to invert the AONT and recover its data. For this reason, the previous solution should be preferred in practice. An example will be given in Section \ref{Performance}.

\section{AONT-LT}
\label{AONT_LTS}

Wishing to pursue the target of data storage in a large number of nodes (greater than that imposed by the size of the RS code) without resorting to fragmentation (the latter having the drawbacks discussed in Section \ref{AONT_RS}) we can resort to a different solution, that is at the basis of our proposal. It consists in replacing the RS code with an LT code, thus realizing an AONT-LT system. A short review of the LT encoding and decoding processes is reported next.

Although non-binary LT codes can be conceived as well \cite{Liva2010}, in this paper we prefer to consider binary LT codes, which is the favorite choice in distributed storage systems because of their simplicity. To obtain a single coded bit:
\begin{enumerate}
\item a degree $d$ is randomly chosen from a probability distribution $p(d)$, that must be properly designed;
\item $d$ input bits are randomly selected from the input word (having length $k$); 
\item a coded bit is calculated as the XOR of the $d$ input bits chosen in step 2.
\end{enumerate}

\begin{figure}[tb]
\centering
\includegraphics[width=3.5in]{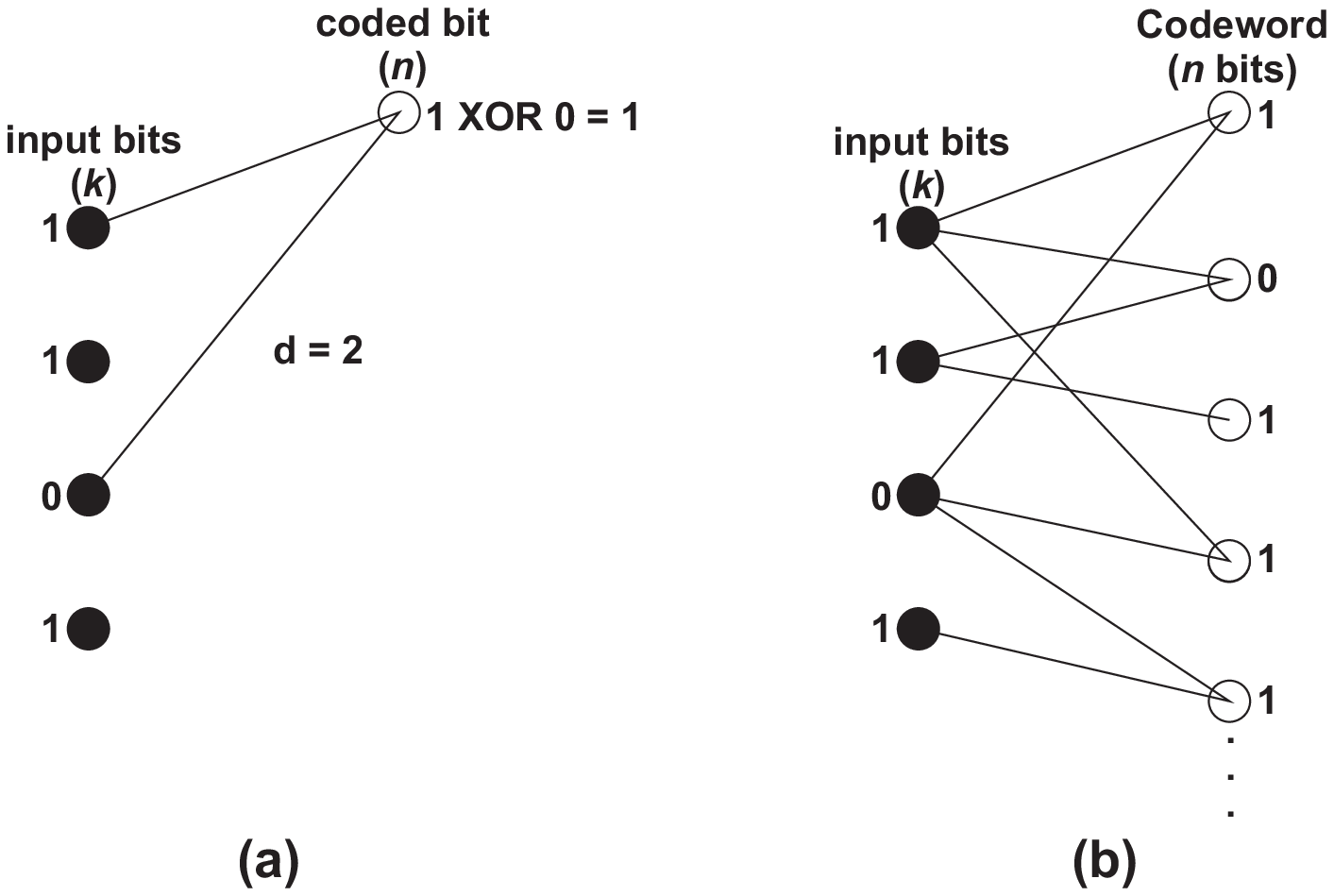}
\caption{Example of (a) LT single bit encoding operation and (b) corresponding LT coding graph.}
\label{fig.LT_coding_graph}
\end{figure}

These 3 steps are repeated $n$ times to generate an LT codeword of $n$ bits. Each encoded bit is independent of the previous encoded bits and this is the reason why the LT codeword length can be freely varied during encoding. The LT encoding process can also be described through a bipartite graph, the latter being particularly useful in the decoding process. Both these representations are shown, for a toy example, in Fig. \ref{fig.LT_coding_graph}.

Looking at the figure, we have also an idea of the very low computational complexity needed to perform the LT encoding process. It reduces to XOR operations between the input bits, and this kind of operations is very fast in modern CPUs.

Different decoding algorithms can be applied to LT codes. Among them, a prominent role is played by message-passing algorithms like BP. The codeword bits have the meaning of check nodes in the bipartite graph. The object of decoding is to recover the input bits on the basis of the parity relationships established by the check nodes. In short, decoding proceeds as follows:

\begin{enumerate}
\item Find a check node with degree one and assign it to the corresponding input symbol. If there is no such a node, halt decoding and report its failure.
\item Add the value of the input symbol disclosed at step 1 to all check nodes connected to it.
\item Remove from the graph all edges connected to the related input symbol. 
\item Iterate the first three steps until all input symbols are recovered.
\end{enumerate}

\begin{figure}[tb]
\centering
\includegraphics[width=3.5in]{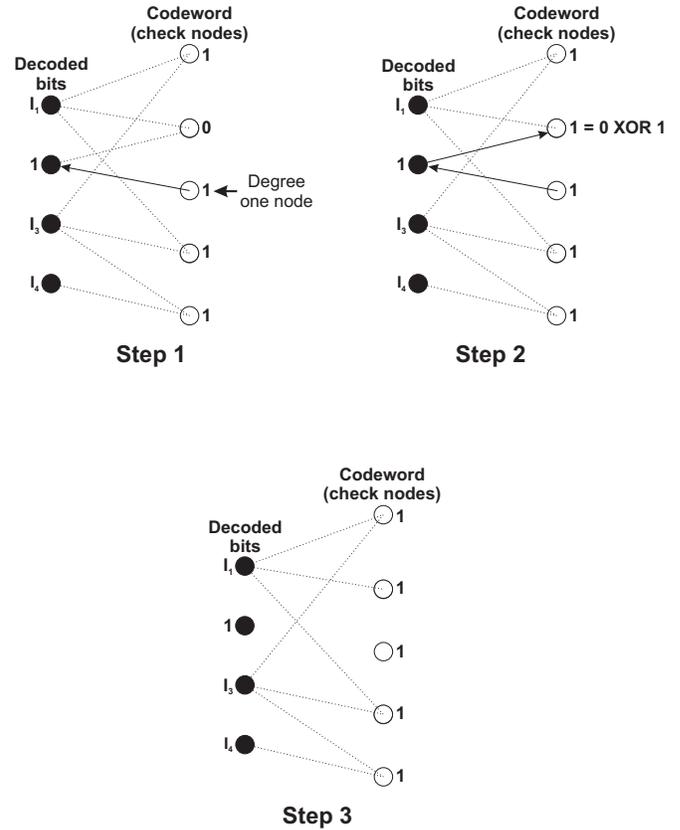}
\caption{First iteration of BP decoding algorithm for the LT graph in Fig. \ref{fig.LT_coding_graph}(b).}
\label{fig.LT_decoding_graph}
\end{figure}

The first iteration of the decoding procedure is shown in Fig. \ref{fig.LT_decoding_graph}, for the same toy example of Fig. \ref{fig.LT_coding_graph}.

Obviously, in the presence of erasures, some check nodes are missing and the reconstruction of the input bits is more difficult. A crucial point concerns the choice of the distribution $p(d)$. Generally speaking, a trade-off should be found between the need to have nodes with low degree (as we have seen, degree one is required to start and feed the decoding process) and the desire to have nodes with high degree, which guarantee widespread connections with the input bits. In any case, the mean value of $d$ should be maintained small, as this ensures a reduced number of operations. 

%We note that we assume $p(d)$ and the coding graph are known at both receiver and transmitter side, and these informations can be exchanged in many different ways as explained in \cite{Luby2002}.
In previous literature, a lot of rules have been presented to design distributions able to comply efficiently with these requirements. Among them, in our analysis and simulations, we have selected the so-called robust soliton distribution (RSD) \cite{Luby2002}.  The robustness of RSD is based on a couple of parameters, $c$ and $\delta$, which can be adjusted to obtain optimal performance with different length \cite{Chen2012}.

\begin{figure}[tb]
\centering
\includegraphics[width=3.5in]{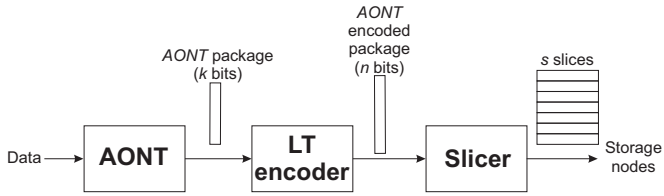}
\caption{AONT-LT principle.}
\label{AONT_LT}
\end{figure}

Wishing to combine the LT code with the AONT function, the system we propose is shown in Fig. \ref{AONT_LT}. In comparison with Fig. \ref{fig.AONT_RS}, we see that the slicer is downstream of the encoder (while it was upstream for the AONT-RS).
The reason for this choice is twofold: on one hand, contrary to RS codes, LT codes have no need of a fixed length at the input; on the other hand, it is known that the performance of LT codes gets better and better for high values of $k$. Based on these facts, it convenient to treat the AONT packet as a whole, acting on the codeword length $n$, variable in its turn ``on the fly'', to reach the desired rate. After encoding, instead, the slicer serves to create $s$ slices, each one composed by $\frac{n}{s}$ bits where $s$ can assume any value; for example we can fix $s$ equal to the number of nodes available in the storage system.

In the next section we will show, through an explicit numerical example, that the scheme in Fig. \ref{AONT_LT} outperforms the AONT-RS variant in Fig. \ref{fig.AONT_tRS}, thus permitting to take advantage of the availability of a large number of storage nodes, without the constraints imposed by the RS encoding scheme.

\section{NUMERICAL RESULTS}
\label{Performance}
As a simple but significant example, let us suppose to have a file with length $k_s = 10$ kB that we wish to store in $n_s = 2540$ available nodes. Using the AONT-RS solution, this implies, for example, the application of $t = 10$ times a RS$(254, 127)$ code (a code of this type is obtained shortening by $1$ the RS$(255, 128)$ code). As regards the AONT-LT solution, instead, though in principle one could use fragmentation as well, we can implement the scheme in Fig. \ref{AONT_LT}, according to which the $10$ kB file is encoded as a whole. The LT code adopted uses an RSD with $c = 0.1$ and $\delta = 0.9$.

\begin{figure}[tb]
\centering
\includegraphics[width=3.5in]{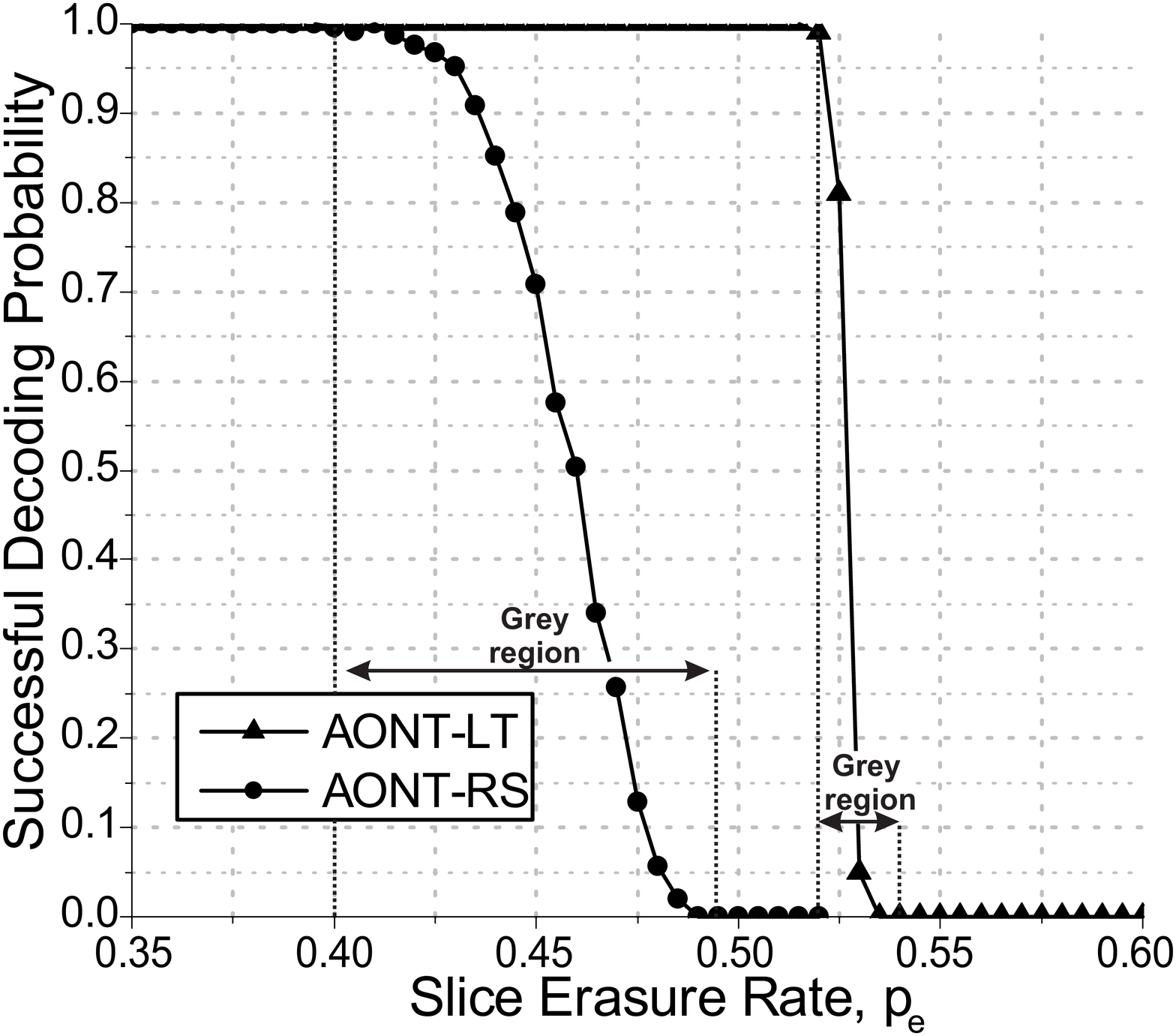}
\caption{Comparison between the successful decoding probabilities of the AONT-LT and the AONT-RS when, for the latter, the AONT is applied to the whole set of $k_s$ slices.}
\label{fig.SDP}
\end{figure}

The performance of the two schemes, in terms of successful decoding probability, is shown in Fig. \ref{fig.SDP}. In order to have a sufficient statistical confidence for the results, for each value of $p_e$ we have considered $150$ simulations.

As mentioned above and expected, because of the application of fragmentation, the curve of the AONT-RS scheme has not a step behavior and a successful decoding probability different from $1$ (perfect reconstruction) and $0$ (null reconstruction) appears. Such an event that, for this particular example, occurs for values of the slice erasure rate roughly comprised between $p_e = 0.4$ and $p_e = 0.49$, defines a \textit{grey region} that is unfavorable in practice and whose extension should be made as small as possible. In that region, in fact, the authorized client has no guarantee that the stored information is correctly retrieved while, at the same time, the attacker can recover some part of the information. The grey region is present for the AONT-LT scheme, too, where it is due to the non-MDS character of the LT code. However, its extension is much smaller being approximately comprised, for the present case, between $p_e = 0.52$ and $p_e = 0.54$.

Incidentally, we also observe that the solution using the LT code has better correction capability; once again, this is a consequence of the fragmentation mechanism where the failure of even 1 decoding out of $t$ inhibits the correct reconstruction because of the action of the AONT. For this specific example, this means that, when using the AONT-RS solution, the client must have access to at least $60\%$ of the nodes to be practically sure to reconstruct the file (versus $50\%$ for the case without fragmentation). On the opposite, when using the AONT-LT solution, the authorized client is practically sure to reconstruct the file even having access to less than $50\%$ of the nodes. This improved correction capability is a positive feature from the client's standpoint, but it is also positive from the attacker's standpoint. However, taking into account that the benchmark is constituted by the step behavior centered at $p_e = 0.5$, we can say that the adoption of the LT code introduces a slight deviation with respect to the ideal case, while permitting to have the advantages related to the large number of storage nodes used.

\begin{figure}[tb]
\centering
\includegraphics[width=3.5in]{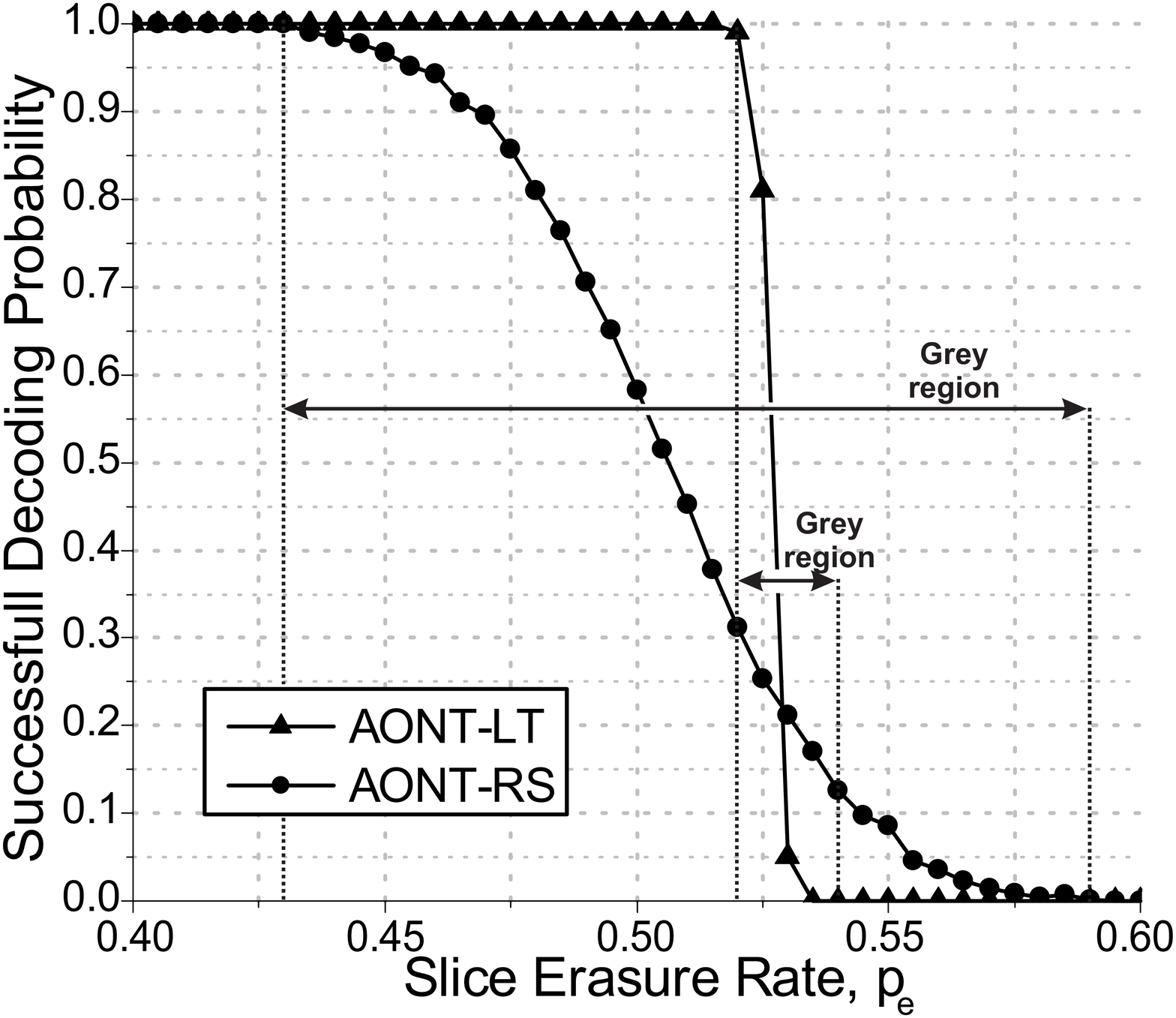}
\caption{Comparison between the successful decoding probabilities of the AONT-LT and the AONT-RS when, for the latter, the AONT is applied to each set of $k$ slices.}
\label{fig.RDR}
\end{figure}

Figure \ref{fig.SDP} refers to the case in which the AONT is applied to the whole set of $k_s$ slices. As mentioned at the end of Section \ref{AONT_RS}, by applying the AONT individually to every block of $k$ slices entering each RS code, the successful decoding probability of the AONT-RS scheme improves as well. This is shown, for the considered example, in Fig. \ref{fig.RDR}. However, the extension of the grey region increases as well, now ranging between $p_e = 0.43$ and $p_e = 0.59$, and globally the curve of successful decoding probability for the AONT-RS scheme becomes worse than that for the AONT-LT schemes from both the client's and the attacker's standpoint. The attacker, in particular, would be able to retrieve a significant amount of information even at a slice erasure rate in the order of $0.54$, where the AONT-LT scheme ensures instead maximum security.

\section{CONCLUSION}
\label{Concl}

In distributed storage systems, and particularly in CSS, there can be a number of advantages in using a large number of storage nodes. On the other hand, when employing the AONT-RS system, the number of nodes is limited by the size of the Galois field over which the code is designed and this size cannot be large in order to limit the co/decoding complexity. The problem can be overcome by using fragmentation, but we have shown that this yields strong deviations of the successful decoding probability from the ideal step behavior which is typical of MDS codes.

So, we have proposed a variant of the joint AONT function plus error correcting code scheme, where instead of using an RS code we have used an LT code. We have verified that this new proposal is able to outperform the AONT-RS scheme with fragmentation, while maintaining a limited extension for the grey region where the system does not work as desired.

The conclusion obtained has been shown looking at a specific example, though randomly chosen. Further work is in progress in view of generalization, also including more significant and practical cases, e.g., as regards the file size and the number of nodes.

\bibliographystyle{IEEEtran}
\bibliography{Archive}

\begin{thebibliography}{10}
\providecommand{\url}[1]{#1}
\csname url@samestyle\endcsname
\providecommand{\newblock}{\relax}
\providecommand{\bibinfo}[2]{#2}
\providecommand{\BIBentrySTDinterwordspacing}{\spaceskip=0pt\relax}
\providecommand{\BIBentryALTinterwordstretchfactor}{4}
\providecommand{\BIBentryALTinterwordspacing}{\spaceskip=\fontdimen2\font plus
\BIBentryALTinterwordstretchfactor\fontdimen3\font minus
  \fontdimen4\font\relax}
\providecommand{\BIBforeignlanguage}[2]{{%
\expandafter\ifx\csname l@#1\endcsname\relax
\typeout{** WARNING: IEEEtran.bst: No hyphenation pattern has been}%
\typeout{** loaded for the language `#1'. Using the pattern for}%
\typeout{** the default language instead.}%
\else
\language=\csname l@#1\endcsname
\fi
#2}}
\providecommand{\BIBdecl}{\relax}
\BIBdecl

\bibitem{Shamir1979}
A.~Shamir, ``How to share a secret,'' \emph{{ACM} Communications}, vol.~22,
  no.~11, pp. 612--613, Nov. 1979.

\bibitem{Rabin1989}
M.~O. Rabin, ``Efficient dispersal of information for security, load balancing,
  and fault tolerance,'' \emph{Journal of the Association for Computing
  Machinery}, vol.~36, no.~2, pp. 335--348, Apr. 1989.

\bibitem{McEliece1981}
R.~McEliece and D.~Sarwate, ``On sharing secrets and {R}eed-{S}olomon code,''
  \emph{{ACM} Communications}, vol.~24, pp. 583--584, Sep. 1981.

\bibitem{Cao2012}
N.~Cao, S.~Yu, Z.~Yang, W.~Lou, and Y.~T. Hou, ``{LT} codes-based secure and
  reliable cloud storage service,'' in \emph{Proc. 31st Annual IEEE
  International Conference on Computer Communications (IEEE INFOCOM 2012)},
  Orlando, FL, Mar. 2012.

\bibitem{Luby2002}
M.~Luby, ``{LT} codes,'' in \emph{Proc. 43rd Ann. IEEE Symposium on Foundations
  of Computer Science}, Vancouver, Canada, Nov. 2002, pp. 271--282.

\bibitem{Rivest1997b}
R.~Rivest, ``All-or-nothing encryption and the package transform,'' in
  \emph{Proc. 4th International Workshop on Fast Software Encryption (FSE
  '97)}, ser. LNCS, E.~Biham, Ed., vol. 1267.\hskip 1em plus 0.5em minus
  0.4em\relax Springer, 1997, pp. 210--218.

\bibitem{Mulazzani2011}
M.~Mulazzani, S.~Schrittwieser, M.~Leithner, M.~Huber, and E.~Weippl, ``Dark
  clouds on the horizon: Using cloud storage as attack vector and online slack
  space,'' in \emph{Proc. 20th USENIX Security Symposium}, San Francisco, USA,
  Aug. 2011.

\bibitem{Drago2012}
I.~Drago, A.~Sperotto, M.~Mellia, R.~Sandre, M.~Munaf{\`o}, and A.~Pras,
  ``Inside dropbox: Understanding personal cloud storage services,'' in
  \emph{Proc. 2012 ACM conference on Internet Measurement Conference}, Boston,
  MA, Nov. 2012.

\bibitem{Resch2011}
J.~Resch and J.~Plank, ``{AONT-RS}: Blending security and performance in
  dispersed storage systems,'' in \emph{Proc. 9th {USENIX} Conference on File
  and Storage Technologies ({FAST})}, San Jose, USA, Feb. 2011.

\bibitem{Liva2010}
G.~Liva, E.~Paolini, and M.~Chiani, ``Performance versus overhead for fountain
  codes over ${F}_q$,'' \emph{{IEEE} Commun. Lett.}, vol.~14, no.~2, pp.
  178--180, Feb. 2010.

\bibitem{Chen2012}
G.~T. Chen, L.~Cao, F.~Zaho, H.~Zheng, and M.~Pan, ``Analysis of robust soliton
  distribution for {LT} code,'' in \emph{Proc. 11th IEEE International
  Conference on Signal Processing (IEEE ICSP 2012)}, vol.~2, Beijing, China,
  Oct. 2012, pp. 1546--1549.

\end{thebibliography}

\end{document}